\documentclass[aps,twocolumn,footinbib,nobalancelastpage,pra,superscriptaddress]{revtex4-2}
\usepackage{amsmath,amssymb}
\usepackage[colorlinks=true,citecolor=blue]{hyperref}
\usepackage{graphicx}
\usepackage{soul}
\usepackage{braket}
\usepackage{float}
\usepackage{amsmath,amsbsy,amsfonts,amssymb,amsxtra}
 
\usepackage[table,xcdraw,dvipsnames]{xcolor}

\begin{document}

%\title{A chaotic model of quantum black holes}

%\title{An information scrambling chiral model of a quantum black hole}

\title{Optimally scrambling chiral spin-chain with effective black hole geometry}

\author{Aiden Daniel}
\thanks{These authors contributed equally}
\affiliation{School of Physics and Astronomy, University of Leeds, Leeds LS2 9JT, United Kingdom}

\author{Andrew Hallam}
\thanks{These authors contributed equally}
\affiliation{School of Physics and Astronomy, University of Leeds, Leeds LS2 9JT, United Kingdom}

\author{Matthew D. Horner}
\affiliation{School of Physics and Astronomy, University of Leeds, Leeds LS2 9JT, United Kingdom}
\affiliation{Aegiq Ltd., Cooper Buildings, Arundel Street, Sheffield, S1 2NS, United Kingdom}

\author{Jiannis K. Pachos}
\affiliation{School of Physics and Astronomy, University of Leeds, Leeds LS2 9JT, United Kingdom}

\begin{abstract}

%Black holes exhibit a complex range of behaviors stemming from their central role in unifying quantum mechanics with general relativity. 
There is currently significant interest in emulating the essential characteristics of black holes, such as their Hawking radiation or their optimal scrambling behavior, using condensed matter models. In this article, we investigate a chiral spin-chain, whose mean field theory effectively captures the behavior of Dirac fermions in the curved spacetime geometry of a black hole. We find that within the region of the chain that describe the interior of the black hole, strong correlations prevail giving rise to many-body chaotic dynamics. Employing out-of-time-order correlations as a diagnostic tool, we numerically compute the associated Lyapunov exponent. Intriguingly, we observe a linear increase in the Lyapunov exponent with temperature within the black hole's interior at low temperatures, indicative of optimal scrambling behavior. This contrasts with the quadratic temperature dependence exhibited by the spin-chain on the region outside the black hole. Our findings contribute to a deeper understanding of the interplay between black hole geometry and quantum chaos, offering insights into fundamental aspects of quantum gravity.

\end{abstract}

\date{\today}
\maketitle{}

{\bf \em Introduction:--}  Black holes pose direct challenges to our understanding of fundamental laws of nature. Central to these open questions is the black hole information paradox, first articulated by Stephen Hawking in the 1970s \cite{Hawking1993,Wald2001}. According to general relativity, the gravitational pull of black holes is so intense that it creates a region known as the event horizon, beyond which information appears to be irretrievably lost. However, the unitarity of quantum mechanics suggests information cannot be destroyed, leading to the question of what happens to the information of an object that falls into a black hole. This apparent contradiction between quantum physics and general relativity has given rise to intense theoretical investigations and remains unresolved to this day.

Recent advancements in theoretical physics have provided new insights into black hole dynamics, particularly through the investigation of quantum information scrambling. Information scrambling refers to the rapid and thorough mixing of information within a quantum system. It is thought that black holes exhibit optimal scrambling behavior, leading to the rapid thermalization of newly engulfed quantum information. Currently, toy models that exhibit maximal scrambling, such as the $(0+1)$D Sachdev-Ye-Kitaev (SYK) model~\cite{Sachdev1993,Standford2016,kitaevtalk2016,Kitaev_2018,polchinski2016spectrum,Fu2016}, are related to $(1+1)$D black holes only through the AdS$_{1+1}$/CFT$_{0+1}$ duality \cite{Jensen2016}, with direct black hole models that exhibit this behavior still lacking.

In this study, we explore the quantum properties of $(1+1)$D black holes using a recently introduced chiral spin-chain model~\cite{horner2023,forbes2024}. The mean field theory limit of this model effectively describes Dirac fermions in a black hole background geometry, which is similar to the semiclassical limit of quantum gravity~\cite{SMMorsink1991}. However, within the region of the chain representing the black hole's interior, the mean field theory description breaks down due to the dominance of strong correlations. The natural question arises if this strongly correlated region gives rise to optimal scrambling behaviour. 

To probe the scrambling behaviour of the chiral spin-chain, we numerically investigate its Out-of-Time-Order correlators (OTOCs). OTOCs are a special class of quantum correlation functions that determine the Lyapunov exponent, capable of diagnosing early-time chaotic behaviour~\cite{Shenker2014,Hosur2016}. Our model exhibits similar scrambling behavior as the SYK model, so we employ the approach presented in~\cite{Kobrin2021} for the numerical analysis of the SYK model.
By focusing on the system's Lyapunov exponent at low temperatures~\cite{Jensen2016,maldacena2016conformal,Engelsy2016, Kitaev_2018} 
we observe that it behaves linearly with temperature at the region of the chain that describes the inside of the black hole, where the chiral interactions are dominant. This signature of optimal scrambling contrasts with the quadratic behaviour observed in the outside region of the black hole, where the chiral interactions have a perturbative effect on top of an XY coupling~\cite{Banerjee2017}. The functional dependence of the Lyapunov exponent on temperature is analysed for various coupling regimes and system sizes showing a robust behaviour and fast convergence to the expected thermodynamic values. Therefore, our chiral model reveals an intricate interplay between black hole geometry and quantum chaos behaviour as expected from a comprehensive quantum gravity description.

\begin{figure}
    \centering
\includegraphics[width=\columnwidth]{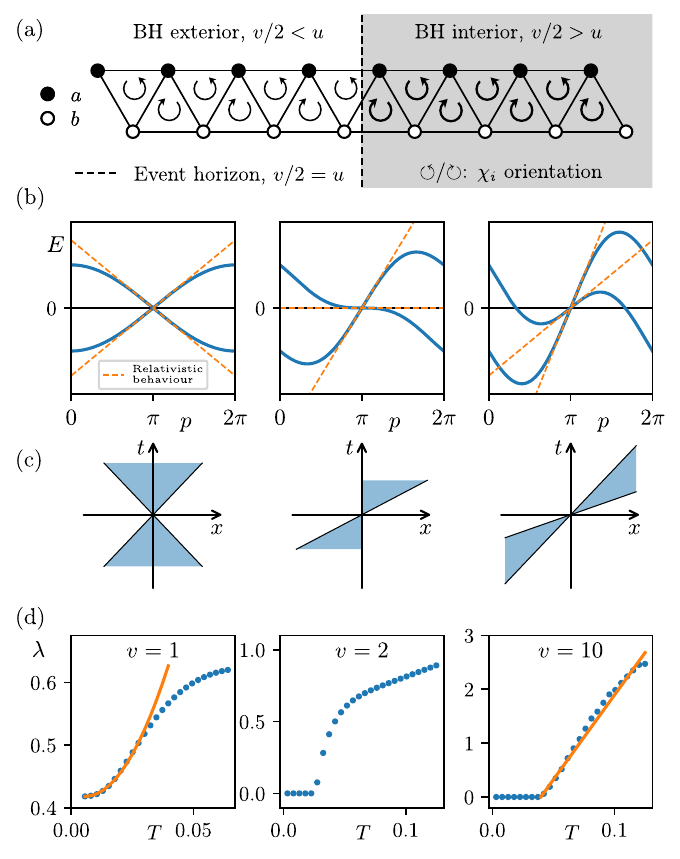}
    \caption{
    (a) The chiral spin-chain \eqref{ham} with position dependent chiral coupling $v$, while $u$ is constant. The chiral interaction, $\chi_i$, acts on three successive spins and has alternating orientation. The fermionic sites $a$ and $b$ represent the unit cell of the mean field theory (MFT), where the spin of the Dirac field is encoded. The MFT gives the black hole (BH) geometry with $v$ small on the left of the chain ($\frac{v}{2} < u$, outside BH) or large on the right ($\frac{v}{2} > u$, inside BH), with the horizon positioned at $\frac{v}{2} = u$. (b) The dispersion relation $E(p)$ obtained from the MFT description of homogeneous periodic chains. In the low energy limit it faithfully reproduces the behaviour of Dirac fermions in black hole geometry given by \eqref{eq:cont_act}. (c) The lightcones, reciprocal to the dispersion relation that describe the effective metric \eqref{eqn:metric} of the system for various values of $v$. (d) The Lyapunov exponent of the fully interacting model exhibits quadratic behaviour $\lambda\propto T^2$ (orange line) on the outside of the BH, where chiral interactions act perturbatively. Maximal scrambling is observed on the inside of the BH with $\lambda\propto T$. The orange line corresponds to the predicted optimal scrambling behaviour with slope $2\pi v/2$. The constant offsets tend to zero with system size. Plots are for $N=8$ and $u=1$.}
    \label{fig:Chain}
\end{figure}

{\bf \em The chiral spin-chain:--} The model, first introduced in Refs.~\cite{horner2023,forbes2024}, is shown in Fig.~\ref{fig:Chain}(a). It describes an interacting chain of spin-$1/2$ particles with Hamiltonian
\begin{equation}
H=\frac{1}{2}\sum_{i=1}^N\left[ -{u} \left( S^x_{i}S^x_{i+1} + S^y_{i}S^y_{i+1} \right )
+\frac{v}{2}\boldsymbol{S}_i \cdot \boldsymbol{S}_{i+1} \times \boldsymbol{S}_{i+2}\right],
\label{ham}
\end{equation}
where $u,v$ are taken here to be positive numbers and $\boldsymbol{S}_i=({\sigma}^x_i/2,{\sigma}^y_i/2,{\sigma}^z_i/2)$ with $\sigma_i^\alpha$ ($\alpha = x,y,z$) is the $\alpha$-Pauli matrix of the $i$th spin. Notably, this is the XY model with an additional three-spin chirality term 
\begin{equation}\label{chiralterm}
    \chi_i=\boldsymbol{S}_i \cdot \boldsymbol{S}_{i+1} \times \boldsymbol{S}_{i+2},
\end{equation}
that introduces interactions~\cite{Pachos05,Tsomokos2008,horner2023,forbes2024}. The enumeration of the sites in the chiral interaction term causes the chirality to alternate along the chain, as shown in Fig.~\ref{fig:Chain}(a). Unless otherwise stated we adopt open boundary conditions.

As $v$ is increased the model undergoes a quantum phase transition from a gapless XY-phase, to a gapless chirally-ordered phase, where the total chirality $\langle \chi \rangle = \sum_i \langle \chi_i \rangle$ acts as an order parameter. The critical point is located at $\frac{v}{2} \approx 1.12u$~\cite{horner2023,forbes2024}. For $\frac{v}{2} \lessapprox 1.12u$, the ground state is in a free XY phase with $\langle \chi \rangle = 0$. Using standard bosonisation techniques~\cite{Giamarchi,Miranda,Sreemayee} we found that the interactions were irrelevant and the low energy physics is described by free fermions with renormalised Fermi velocities. On the other hand, for $\frac{v}{2} \gtrapprox 1.12u$ the chiral interaction dominates the XY term and the model transitions to a chiral phase with $\langle \chi \rangle \neq 0$. In this phase bosonisation is more complicated due to the system possessing two Fermi points. We find that in this phase, the model does not remain at the free fermion point, revealing the dominance of the interactions~\cite{forbes2024}.

{\bf \em Black hole background geometry:--} Intriguingly, the model has a geometric interpretation in terms of black hole physics~\cite{horner2023}. We first apply the Jordan-Wigner transformation that maps the spins into fermions. The system then has a unit cell of two fermion sites, $a$ and $b$, sitting at opposite ranks of the triangular ladder shown in Fig.~\ref{fig:Chain}. By employing self-consistent mean field theory (MFT), one can map the interacting spin model to a model of free fermions on a lattice. We can investigate the behaviour of the model by taking homogeneous coupling $v$ and adopting periodic boundary conditions to extract the dispersion relation. This description faithfully captures the phase diagram of the model, albeit with critical point at $\frac{v}{2} = u$. The dispersion relation of the model, shown in Fig.~\ref{fig:Chain}(b), at low energies, i.e., in the continuum limit, can be faithfully reproduced by the Dirac action on a fixed curved spacetime background
\begin{equation}
S_{\rm{MFT}}  =  \int \mathrm{d}^{1+1}x |e| \bar{\psi}(x)\left( ie_a^{\ \mu} \gamma^a \overset{\leftrightarrow}{\partial_\mu} \right) \psi(x), \label{eq:cont_act}
\end{equation}
where $a=0,1$ are local inertial frame indices, $\mu=t,x$ are coordinate indices; the spinor field is given in terms of the unit cell fermions as $\psi(x)=(a(x),b(x))^T/\sqrt{|e|}$ as shown in Fig.~\ref{fig:Chain}(a); $A\overset{\leftrightarrow}{\partial_\mu}B =\frac{1}{2} \left( A \partial_\mu B - (\partial_\mu A)B \right)$; $\gamma^a = (\sigma^z,-i\sigma^x)$; and $|e|=\text{det}(e^a_{\ \mu})$. The zweibein $e_a^{\ \mu}$ are related to the spacetime metric by $g_{\mu\nu} = e^a_{\ \mu}e^b_{\ \nu}\eta_{ab}$, where $\eta_{ab} = \text{diag}(1,-1)$ is the Minkowski metric and $g_{\mu\nu}$ is given by~\cite{horner2023}
\begin{equation}
\mathrm{d}s^2 = \left( 1 - \frac{v^2}{4u^2} \right)\mathrm{d}t^2 - \frac{4v}{u^2}\mathrm{d}t \mathrm{d}x - \frac{16}{u^2}\mathrm{d}x^2.
\label{eqn:metric}
\end{equation}
This is the Schwarzschild metric of a black hole expressed in Gullstrand-Painlev\'e coordinates~\cite{Volovik2003,Volovik1,Volovik2,Volovik3} (see SM for more details), which has also been observed in other synthetic black hole systems~\cite{Beule2021,Morice2021,Konye2022,Haller2023}. 

We now take the coupling $v$ to be a function of position, $v(x)$, varying monotonically from small to large values. If it is slowly-varying compared to the lattice spacing, then the continuum description in terms of the Dirac equation remains valid. In this case the event horizon is located at $\frac{v}{2} = u$, where $\frac{v}{2} < u$ corresponds to the outside of the black hole and $\frac{v}{2} > u$ corresponds to the inside, as shown in Fig.~\ref{fig:Chain}(c). In Gullstrand-Painlev\'e coordinates the light cone tilts when approaching the black hole, having eventually both light directions pointing towards its centre inside the black hole, as shown in Fig.~\ref{fig:Chain}(c) (Right). We see that the event horizon aligns well with the boundary between the two phases of the spin-chain, where the chiral phase is inside the black hole and the XY phase outside. Using the mean field description it was shown in Ref.~\cite{horner2023} that a free particle that passes through the phase boundary of our model emerges as a thermal radiation with the Hawking temperature, similar to other models~\cite{Wilczek,Yang,Sabsovich,PhysRevB.109.014309,PhysRevResearch.4.043084,Stone_2013,Steinhauer_2016,Kosior,Roldan-Molina,Rodriguez,Retzker,Guan,Hang,Huang}. We expect the thermalisation to Hawking temperature to be valid in the fully interacting model as the MFT is still valid around the horizon, only breaking down deep inside the black hole where the chiral interactions are dominant.

Apart from reproducing the semiclassical behaviour of a black hole, our model also exhibits a chaotic behaviour, as we shall see in the following. This can be quantified by the Lyapunov exponent shown in Fig.~\ref{fig:Chain}(d). For small $v$ we obtain a Lyapunov exponent $\lambda\propto T^2$, as expected from perturbative interactions~\cite{Kim2020}. For large $v$, i.e. inside the black hole, the numerically obtained Lyapunov exponent exhibits linear behaviour $\lambda \propto T$ \cite{Maldacena_2016}, with a slope which is in excellent agreement with the predicted optimal scrambling behaviour. The constant offsets in Fig.~\ref{fig:Chain}(d) for $v=1$ and $v=10$ go to zero with system size, as we shall see in the following. 

{\bf \em Quantum chaos inside black hole:--} A natural question to ask is if this black hole geometric analogy of Eq.~\eqref{ham} extends to the thermalising dynamics expected at the interior of a black hole. In particular, we investigate whether the spin model is chaotic for $\frac{v}{2} > u$ (with homogeneous $v$ along the chain) and, more importantly, whether it exhibits maximal information scrambling as expected of a black hole.

\begin{figure}[tb]
    \centering
    \includegraphics[width=\columnwidth]{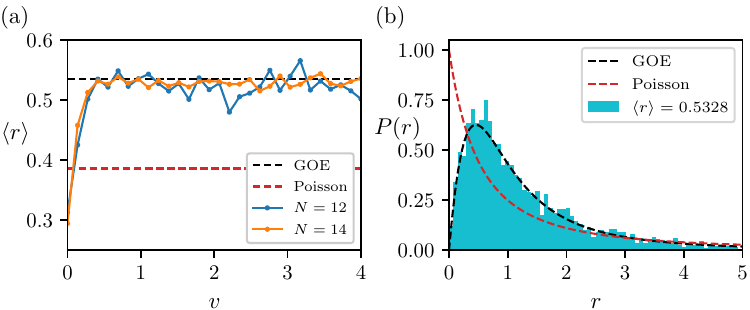}
    \caption{ (a) The average $r$-value~\eqref{rval} of the eigenspectrum of Hamiltonian~\eqref{ham} for different values of $v$ and $N$ ($u=1$). (b) The probability distribution of all r-values in the reduced symmetry sector at $N=20,v=4,u=1$. In both cases we see the model matches the Gaussian orthogonal ensemble (GOE) indicating a non-integrable model~\cite{Atas2013}. Results were computed in the $k=0,z=0, x=+1$ symmetry sector.}
    \label{fig:r_valstats}
\end{figure}

One of the most effective methods for diagnosing the chaotic behaviour of a many-body quantum system is to study its energy level statistics, provided all relevant symmetries have been resolved. We consider the chain with periodic boundary conditions which has translational symmetry, U$(1)$ symmetry and global spin flip symmetry $X=\prod_i\sigma^x_i$. We restrict to the symmetry sector with quantum numbers
$k=0,z=0,x=+1$ of these symmetries, respectively, and determine the  eigenvalues, $\left \{ E_n \right \}$ of \eqref{ham}. We then take the set $\left \{ s_n \right \}$, where $s_n=E_n-E_{n-1}$, and evaluate \cite{Oganesyan2007}
\begin{equation}\label{rval}
    r_n=\frac{\min\{s_n,s_{n-1}\}}{\max\{s_n,s_{n-1}\}}.
\end{equation}
The average of this value $\langle r \rangle$ and the probability distribution over all $r_n$, are shown in Fig.~\ref{fig:r_valstats}. For $v\neq0$, we find Wigner-Dyson statistics indicating that this model is chaotic, a characteristic that becomes more prominent with system size. Notably, we find $\langle r \rangle \approx 0.53$ \cite{Atas2013}, which corresponds to the GOE ensemble. This is perhaps unexpected since $H$ possesses complex matrix elements. Despite this, the model retains time-reversal symmetry due to satisfying the relation $PH^TP=H$ where $P$ is the parity operator, reminiscent of Ref.~\cite{Regnault2016}. We also calculated the Spectral Form Factor which further probes the spectrum for evidence of chaotic behaviour but did not observable any notable additional behavior \cite{Liu2018,Gaikwad2019,Cotler2017}.

{\bf \em Lyapunov exponent and optimal scrambling:--} The average level spacing $\langle  r \rangle$ is a crude measure of the chaotic behaviour of a quantum system, in the thermodynamic limit we expect Wigner-Dyson statistics for all models except fine-tuned integrable systems. Therefore, we need a more precise measure of the chaotic behaviour of the system inside the black hole to determine whether the scrambling of quantum information is optimal, as it is the case for the SYK model~\cite{Kobrin2021}. 

As a diagnostic tool we will employ the Lyapunov exponent, $\lambda$, that quantifies the rate of thermalisation of a chaotic system~\cite{Hallam_2019}. In the quantum mechanical framework, $\lambda$ is calculated using the decay in out-of-time-order correlators (OTOCs) with respect to some local operator $O_i$ on site $i$, as a function of time, $t$. We primarily focus on the regularised OTOC
\begin{equation}\label{otocs}
    C(t)=\langle O_i(t)\rho^{1/4}O_j(0)\rho^{1/4}O_i(t)\rho^{1/4}O_j(0)\rho^{1/4} \rangle,
\end{equation}
where $\rho = \exp(-\beta H)/\mathcal{Z}$, with the partition function $\mathcal{Z}=\textup{Tr} \exp(-\beta H)$ and $\beta=1/T$ is the inverse temperature. We also scale such that $C(0)=1$. 

The regularised version of OTOCs is suitable for investigating small temperature behaviours and exhibits fast convergence even for small system sizes~\cite{Kobrin2021}. This should be contrasted to the unregularized correlator which is subject to stronger finite-size corrections at low temperatures~\cite{Lantagne2020}. Using the regularised version, $C(t)$, we are restricted to exact diagonalisation techniques. Further restrictions are placed on system size $N\leq 13$ due to the need of time evolution. Fortunately, due to the fast convergence of our model with system size, we find this to be sufficient for our study. Unless otherwise stated, we take $O_i=S_{N/2}^x$ and  $O_j=S_{N/2-2}^x$ for $N$ even and $O_i=S_{(N+1)/2}^x$, $O_j=S_{(N-3)/2}^x$, for $N$ odd.
While the choice of $O=S^x$ restricts the use of system's symmetries, it is reminiscent of the Majorana fermionic operators in the SYK model and successfully unearths the desired optimality behavior. We show below that the regularised correlators we choose, allow us to faithfully extract the chaotic behaviour of the model even with moderate system sizes.

With maximally quantum information scrambling models, one expects an exponential decay in the OTOCs defined in Eq.~\eqref{otocs} with an associated Lyapunov exponent, $\lambda$, indicating the rate of decay. The numerical recipe we employ here is identical to that presented in Ref.~\cite{Kobrin2021} in the context of SYK. We find that the same method is effective at identifying the scrambling behaviour of our model. Mirroring this study of the SYK model, we fit the numerical data of Eq. \eqref{otocs} to the semiclassical function of the OTOC at low temperatures
\begin{equation}\label{otocfit}
    C(t)= U\Big(\frac{1}{2},1,Ne^{-    \lambda t}\Big) \sqrt{N} e^{-\lambda t/2},
\end{equation}
where $U$ is the Kummer's confluent hypergeometric function and $\lambda$ is the fitted Lyapunov exponent. In general, we expect $\lambda$ to depend on the coupling $v$ and the temperature $T$ of the model, while we keep $u=1$.
\begin{figure}
    \centering
 \includegraphics[width=\columnwidth]{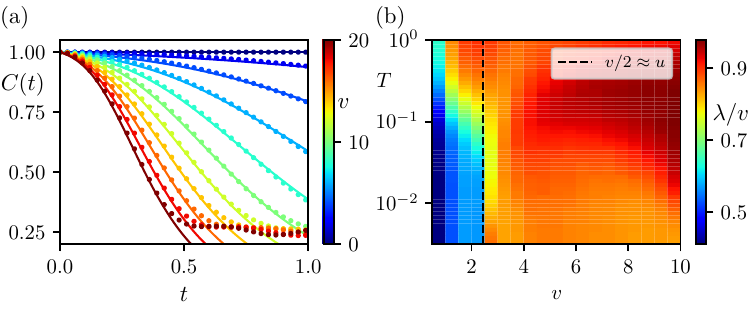}
    \caption{Out-of-time-ordered correlators, $C(t)$, and Lyapunov exponent, $\lambda$, of the chiral spin-chain for various coupling regimes. (a) Coloured dots show the numerically evaluated $C(t)$ given in Eq. \eqref{otocs} for various values of $v$ (shown in the color bar), and $T=\infty$. Lines show the fit of Eq.~\eqref{otocfit} with $\lambda=0.78v$, which improves for $v$ deep in the chiral phase.
    (b) Using the same process and parameters as in (a), we compute the OTOCs, while varying both $T$ and $v$ and extract $\lambda$ via fitting~\eqref{otocfit}. Large values of $\lambda$ are observed for large $v$ (chiral regime) and large temperatures $T$ that probe the full spectrum of the Hamiltonian. Black dashed line indicates the phase transition at ${v}/{2}\approx u$, where a clear change in behaviour at low temperatures is witnessed. Both plots are computed with $N=10$, $u=1$.}
    \label{fig:enter-label}
\end{figure}
Fig.~\ref{fig:enter-label} (a) shows the behaviour of OTOCs for various values of coupling $v$ when $u=1$ and the system size is $N=10$. We verify that the OTOCs exponentially 
decrease in time as seen by the fit (solid lines) in Fig.~\ref{fig:enter-label} (a) with $\lambda=0.78v$. This exponential scrambling behaviour is present for large $v$, that take the spin-chain in the chiral regime, i.e. inside the black hole. These results remain largely unchanged as we vary the system size demonstrating the fast convergence in the properties of the chiral spin-chain. 

With the confirmed presence of exponential decrease in the OTOCs with time, we have already substantiated the argument for the presence of exponential scrambling characterized by the Lyapunov exponent. We now investigate the change in the Lyapunov exponent with temperature where temperature defines the average energy of the density matrix. Zero temperature corresponds to the ground state while infinite temperature corresponds to a uniform superposition of all eigenstates. Due to expected thermal behaviour of mid spectrum eigenstates in chaotic models, one expects the Lyapunov exponent, $\lambda$, to increase with $T$ to some maximal bound. The behaviour of $\lambda$ as a function of the coupling $v$ and the temperature $T$ is given in Fig.~\ref{fig:enter-label} (b), where the increase in $\lambda$ is observed with $v$ and $T$, as expected. We also note a sudden change in the behaviour of $\lambda/v$ as $v$ crosses the phase transition point $\frac{v}{2}\approx u$ at small temperatures, revealing the corresponding dramatic change in the scrambling behaviour of the model.

\begin{figure}
    \centering
    \includegraphics[width=\columnwidth]{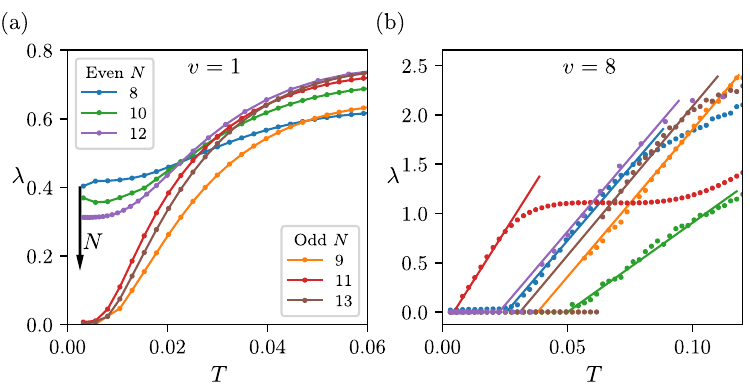}
    \vspace{0mm} % Remove vertical space between the images
    \includegraphics[width=\columnwidth]{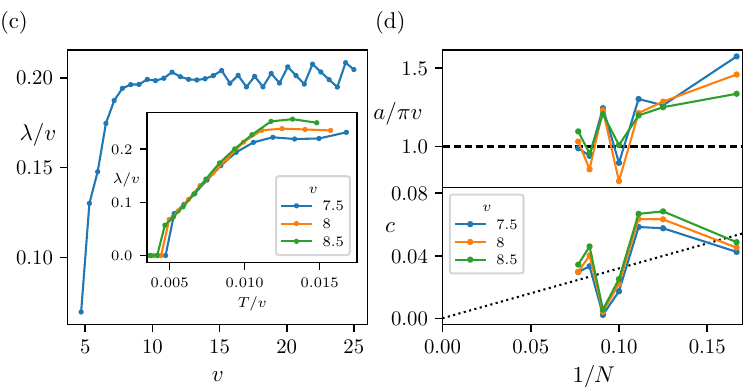}
    \caption{The Lyapunov exponent of the chiral spin-chain simulating the black hole. (a) Outside the black hole ($v=1$ and $u=1$) the Lyapunov exponent exhibits at low temperatures a quadratic behavior. The observed even/odd system size dependence is decreased when $N$ increases. (b) Inside the black hole ($v=8$ and $u=1$) a linear behaviour in the Lyapunov exponent is observed, indicating optimal scrambling. (c) The functional dependence of the linear regime of $\lambda$ on the coupling $v$. When scaling both $\lambda$ and $T$ by $v$ we find that with increasing $v$, the scaled Lyapunov exponent eventually flattens out ($N=8$). (Inset) Upon scaling both $\lambda$ and $T$ axis by $v$, the linear plots collapse on top of each other ($N=12$).
    (d) In the linear regime we fit $\lambda =a(T - c)$ as shown in (b) and extract the slope $a$ and the offset $c$. (d, Top) We plot $a$ scaled by $v/2$ and by $2\pi$ against $1/N$ ($N$ from 6 to 13). We find the slope $\lambda/(v/2)$ tends towards $2\pi$ (dashed line) as $N$ increases. (d, Bottom) The constant offset, $c$, is also extracted. The linear fit to the data (dashed line) showing $c$ tending to 0 in the large $N$ limit.}
    \label{fig:Lyap}
\end{figure}

It is predicted that the Lyapunov exponent, describing the scrambling behaviour in quantum systems, satisfies the universal bound $\lambda/J \leq 2\pi T$, where $J$ is a characteristic coupling of the system~\cite{Maldacena_2016}.
The quantum gravity description of black holes and their holographic duals, such as the SYK model, are known to saturate this bound~\cite{shenker2015stringy}, when $\lambda$ is normalised by appropriately chosen coupling $J$. This optimal scrambling behaviour is analytically and numerically identified in the SYK model for low temperatures~\cite{Maldacena_2016}, and has been experimentally investigated~\cite{Tian2022}. We will now investigate whether the chiral spin-chain is optimally scrambling or not. To achieve that we will quantitatively determine the functional dependence of the Lyapunov exponent on temperature, $T$, for various regimes of $v$ that correspond to the inside and outside of the black hole. Again, we employ the same method seen in \cite{Kobrin2021} when fitting with Eq.~\eqref{otocfit}.

In the low temperature limit we expect that in the weakly-interacting regime, described by $\frac{v}{2}<u$, the Lyapunov exponent will grow quadratically with temperature. In Fig.~\ref{fig:Lyap}(a) we observe that the Lyapunov exponent obtains the quadratic behaviours as $T$ goes to zero, when $v$ is small, i.e. outside the black hole. Note that odd system sizes have a zero offset, i.e. $\lambda\to 0$ as $T\to 0$, while for even $N$ there is a non-zero offset. This offset is a finite system size effect that tends to zero as $N$ increases. 

If the strongly interacting regime, $\frac{v}{2}>u$, exhibits black hole phenomena, then optimal scrambling is expected, witnessed by a linear growth of the Lyapunov exponent, $\lambda/J = 2\pi T$. In Fig.~\ref{fig:Lyap}(b) we observe that, similar to the SYK model~\cite{Kobrin2021}, the numerically obtained Lyapunov exponent has a liner dependence on temperature. Hence, we anticipate that the chiral spin-chain exhibits optimal scrambling at the coupling regime that describes the inside of a black hole. To quantify how accurately the quadratic and linear behaviors are manifested in our system we module a fit in the data of the form $\lambda = a(T^b - c)$ and study $b$ with system size. We find that $b$ takes values 1 and 2 in the corresponding regimes to a very good accuracy (see SM for more details). Furthermore, Fig.~\ref{fig:Lyap}(c) shows that for large enough $v$, the Lyapunov exponent remains more or less constant taking the value $\lambda/v\approx 0.20$. This should be contrasted to other sub-optimal models with a rate of chaos that is parametrically slower than the SYK model~\cite{Stanford_2016,Chowdhury_2017}.

We next investigate the slope of the linear behaviour exhibited by the Lyapunov exponent when $v$ is large. In Fig.~\ref{fig:Lyap}(b) we identify the linear behavior $\lambda = a(T-c)$ for a range of temperatures between $T_{\rm{min}}$ and $T_{\rm{max}}$. Here, $T_{\rm{min}}$ depends on the discreteness of the finite system and tends to zero as $N$ increases, while $T_{\rm{max}}$ depends on the rest of the dynamics of the chiral model.
The saturation of the scrambling bound is achieved for a linear gradient $2\pi$ normalised by the coupling of interactions, given in Hamiltonian \eqref{ham} by $v/2$. In Fig.~\ref{fig:Lyap}(d) (Top) we see that the slope of $\lambda/(v/2)$ as a function of temperature, $2a/v$, tends to $2\pi$ with increasing system size as expected from the optimal scrambling behaviour. Moreover, we observe that the constant offset, $c$, tends to zero with increasing system size, $N$, where the dashed line in Fig.~\ref{fig:Lyap}(d) (Bottom) is a linear fit to the data. Note that both $a$ and $c$ show a strong oscillatory behaviour as a function of $N$, indicating the significance of the boundary effects for the system sizes we considered. Hence, in the strong chiral regime of our simulator, i.e. inside the black hole, we expect to have $\frac{\lambda}{v/2} = 2\pi T$ in the thermodynamic limit. Although, resolving the ambiguity associated with choosing the appropriate energy scale of the model, $J=v/2$, needs a theoretical investigation that goes beyond the scope of this article, we postulate that this expression corresponds to optimal scrambling.

Our numerical investigation shows that as the coupling $v$ varies from small to large values the spin-chain \eqref{ham} undergoes a quantum phase transition. This transition does not only changes its ground state properties from non-chiral to chiral~\cite{horner2023,forbes2024}, but its thermalisation properties change from weakly scrambling to optimal scrambling at $\frac{v}{2}\approx u$, much in the same way as in~\cite{Banerjee2017}. Notably, our system does not have random all-to-all interactions as it is the case for the SYK model~\cite{Kobrin2021}. This locality and uniformity facilitates the fast convergence of our numerical simulations with system size to the expected behaviour. This is manifestly seen in Fig.~\ref{fig:Lyap}, where system size effects become less pronounced with increasing $N$.

{\bf \em Conclusions:--} Our investigation establishes the connection between quantum information scrambling and the behavior of chiral spin chains that encode black hole spacetime geometry. Through numerical analysis of out-of-time-order correlators, we have provided compelling evidence that at the coupling regime representing the interior of the black hole information encoded in the chain scrambles at an optimal rate.

Our findings open avenues for further exploration in several directions. 
%{\color{red} Firstly, unlike the SYK model, our model is not analytically tractable. Nevertheless, it would be advantageous to find more analytical arguments for the optimal scrambling behaviour.} 
Firstly, a theoretical analysis to determine the Lyapunov exponent of our chiral spin chain would complement our numerical findings, offering a deeper understanding of its chaotic behavior.%} 
Additionally, investigating the quantum phase transition at $\frac{v}{2} \approx u$, where both the ground state and scrambling behavior undergo significant changes, presents an intriguing topic for future research. Moreover, this initial investigation lends itself to generalisations to $(2+1)$ or $(3+1)$ dimensional black holes following the methodology presented in~\cite{Aydin}.

Importantly, unlike the SYK model, our model incorporates uniform local chiral interactions, making it more experimentally accessible. Previous studies have shown the feasibility of realizing chiral interactions in optical lattice systems~\cite{Cruz2005,Tsomokos2008}. This suggests the potential for experimental verification of optimal scrambling behavior in laboratory settings. The plausible experimental accessibility of our model opens new possibilities for studying quantum gravity-inspired phenomena in controlled environments.

{\bf \em Acknowledgements:--} We would like to thank Zlatko Papi\'c and Cristian Voinea for inspiring conversations. This work was supported by EPSRC with Grant No. EP/R020612/1 and by the Leverhulme Trust Research Leadership Award RL-2019-015.

\bibliography{ref}

\newpage

\appendix

\section{Lattice representation of Dirac field in black hole background}
\subsection{Mean field approximation}

The system we investigate in the main text is the one-dimensional spin-$\frac{1}{2}$ chain with the Hamiltonian
\begin{equation}
H=\frac{1}{2}\sum_n \left[ -{u} \left( S^x_{n}S^x_{n+1} + S^y_{n}S^y_{n+1} \right )
+\frac{v}{2}\chi_n \right], \label{eq:app_spin_ham}
\end{equation}
where the spin chirality operator is given by
 \begin{equation}
\chi_n  = \boldsymbol{S}_n \cdot \left(\boldsymbol{S}_{n+1} \times \boldsymbol{S}_{n+2}\right),
\label{eq:C4_chiral_op}
\end{equation}
where $\boldsymbol{S}_n = \frac{1}{2}(\sigma^x_n, \sigma^y_n, \sigma^z_n)$ is the spin vector, where $\sigma_n^a$ is the $a$-Pauli matrix acting on the $n$th lattice site for $a \in \{ x,y,z \}$, and the $u,v \in \mathbb{R}$ are couplings with dimensions of energy. In this supplemental material, we assume we are working in the thermodynamic limit.

First we transform from spin operators to Pauli operators. In terms of the Pauli operators, the Hamiltonian is given by
\begin{equation}
\begin{aligned}
H =\sum_n \Big[ & -\frac{u}{8} \left( \sigma^x_{n}\sigma^x_{n+1} + \sigma^y_{n}\sigma^y_{n+1} \right ) \\
&  +\frac{v}{32} \epsilon_{abc} \sigma^a_n  \sigma^b_{n+1} \sigma^c_{n+2}  \Big]
\end{aligned}
\end{equation}
where the repeated Latin indices $a,b,c$ are summed over in the chirality term. We now introduce the ladder operators $\sigma_n^\pm = (\sigma^x_n \pm i \sigma^y_n)/2$ and the Jordan-Wigner transformation defined by~\cite{Coleman_2015}
\begin{align}
\sigma^+_n & = \exp \left( - i \pi \sum_{m < n} c_m^\dagger c_m \right) c^\dagger_n \\
\sigma^-_n & = \exp \left( i \pi \sum_{m < n} c_m^\dagger c_m \right) c_n \\
\sigma^z_n & = 2c^\dagger_n c_n - 1 \label{eq:JW_Z}
\end{align}
where  $c_n$ are a set of fermionic modes obeying the anti-commutation relations $\{ c_n , c_m \} = \{ c_m^\dagger , c_n^\dagger \} = 0$ and $\{ c_n , c_m^\dagger \} = \delta_{mn}$. 
After expressing the Hamiltonian in terms of $\sigma_n^\pm$ and $\sigma^z_n$ and then applying the Jordan-Wigner transformation, we arrive at
\begin{equation}
\begin{split}
H  =   \frac{1}{4} \sum_n \bigg[ & - u c_n^\dagger c_{n+1} - \frac{iv}{4} c^\dagger_n c_{n+2}  \\
& + \frac{iv}{4}  \left( c_n^\dagger c_{n+1} \sigma^z_{n+2} + c^\dagger_{n+1} c_{n+2} \sigma^z_n \right) \bigg] + \text{H.c.}, \label{eq:app_h_interacting} 
\end{split}
\end{equation} 
where for convenience we have left $\sigma^z_n$ alone under the assumption that it represents the Jordan-Wigner transformation of Eq.~~(\ref{eq:JW_Z}).  We see that the model is intrinsically interacting as the fermionic Hamiltonian contains quartic terms which arise from terms like $c_n^\dagger c_{n+1}\sigma^z_{n+2}$ after explicitly substituting in Eq.~\eqref{eq:JW_Z}.

To analyse the behaviour of the interacting model, we apply mean field theory (MFT) to transform the Hamiltonian into an effective quadratic Hamiltonian which can be analytically diagonalised. MFT defines the fluctuation of an operator $A$ as $\delta A = A - \langle A \rangle $, where $\langle A \rangle$ is the expectation value of the operator $A$ with respect to the mean field ground state $|\Omega\rangle$. For a product of two operators we have
\begin{equation}
AB   = \langle A \rangle B + A \langle B \rangle - \langle A \rangle \langle B \rangle + \delta A \delta B,
\end{equation}
where the second order in fluctuations can be ignored. Applying this to the interacting terms of Eq.~\eqref{eq:app_h_interacting} where we always consider $\sigma^z_n$ as one of the operators in the product of Eq.~\eqref{eq:app_h_interacting}, so replace $\sigma^z_n \rightarrow \langle \sigma^z_n \rangle \equiv Z$ and $c_n^\dagger c_{n+1} \rightarrow \langle c_n^\dagger c_{n+1}\rangle \equiv \alpha$, where we have assumed translational invariance. These expectation values are done with respect to the ground state of the mean field Hamiltonian. The Hamiltonian becomes
\begin{equation}
\begin{split}
H_\mathrm{MF}(\alpha,Z) & = \frac{1}{4} \sum_n \bigg[ -\left(u - \frac{iv}{2} Z \right) c^\dagger_n c_{n+1} \\ & - \frac{iv}{4} c^\dagger_n c_{n+2} 
+ 4\mu c^\dagger_n c_n \bigg] + E_0 + \mathrm{H.c.}, \label{eq:H_MF_parametrised}
\end{split}
\end{equation}
where $\mu =  v \mathrm{Im}(\alpha)/4$ is an effective chemical potential controlling the number of particles in the ground state, $E_0 =  v ( Z - 1) \mathrm{Im}(\alpha)/8 $ is a constant energy shift.

Let $|\Omega(\alpha,Z)\rangle$ be the ground state of the Hamiltonian of Eq.~\eqref{eq:H_MF_parametrised}. Self consistency requires
\begin{align}
\langle \Omega(\alpha,Z)| \sigma^z_n | \Omega(\alpha,Z) \rangle & = Z \\
\langle \Omega(\alpha,Z) | c^\dagger_n c_{n+1} |\Omega(\alpha,Z) \rangle & = \alpha
\end{align}
for all $n$. While these two equations have many solutions, we can single one out on physical grounds: the fully interacting Hamiltonian of Eq.~\eqref{eq:app_h_interacting} has particle-hole symmetry, $[H,U]= 0$, where $U$ is the particle-hole transformation with $U c_n U^\dagger = (-1)^n c_n^\dagger$ and $U c_n^\dagger U^\dagger = (-1)^n c_n$. This symmetry implies that $\langle c_n^\dagger c_{n}\rangle = 1/2$ and $\langle c_n^\dagger c_{n+1} \rangle \in \mathbb{R}$ in the ground state. If we require the MFT to retain the particle-hole symmetry, then these conditions imply that $Z = \mathrm{Im}(\alpha) = 0$, and the MFT Hamiltonian becomes
\begin{equation}
H_\text{MF} = \frac{1}{4}\sum_n \left( -u c^\dagger_n c_{n+1} - \frac{iv}{4} c^\dagger_n c_{n+2} \right) + \text{H.c.}. \label{eq:mf_ham}
\end{equation} 
It was shown in Ref.~\cite{horner2023} that this mean field limit faithfully describes the second order phase transition exhibited by the full spin model.

\subsection{The Dirac equation on curved spacetime}

We now briefly introduce the Dirac field on a curved spacetime which we shall use in the next section of the supplemental material. Suppose we have an $N+1$-dimensional spacetime with metric $g_{\mu \nu}$ and a set of veilbein $\{ e_a^{\ \mu} \}$ and their inverses $\{ e_a^{\ \mu} \}$ which are related to the metric via
\begin{equation}
g_{\mu \nu} = e^a_{\ \mu} e^b_{\ \nu} \eta_{ab}, \quad g^{\mu \nu} = e_a^{\ \mu} e_b^{\ \nu} \eta^{ab} ,
\end{equation}
where $\eta_{ab} = \mathrm{diag}(1,-1,-1,\ldots,-1)$ is the Minkowski metric. The veilbein and their inverses also obey
\begin{equation}
    e_a^{\ \mu} e^b_{\ \mu} = \delta^a_b, \quad e_a^{\ \mu} e^a_{\ \nu} = \delta^\mu_\nu.
\end{equation}
 The veilbein $\{ e_a^{\ \mu} \} $ are a set of vector fields that form an orthonormal basis at every point in some patch of $M$. The Latin indices $a,b =0,1,\ldots$ refer to the orthonormal frame indices, whilst the Greek indices $\mu, \nu = t,x,\ldots$ refer to the coordinate indices.

Using this, we can introduce spinor fields on a curved spacetime as a field on $M$ which transforms as a spinor under local Lorentz transformations (transformations that act on Latin indices) and as a scalar under coordinate transformations (transformations that act on the Greek indices). The locally Lorentz invariant and coordinate invariant action for spinor field $\psi$ of mass $m$ on an $N+1$ dimensional spacetime $M$ with metric $g_{\mu \nu}$ is given by~\cite{Nakahara}
\begin{equation}
\begin{aligned}
S & = \frac{i}{2} \int_M \mathrm{d}^{N+1}x |e| \left( \bar{\psi} \gamma^\mu D_\mu \psi - \overline{D_\mu \psi}\gamma^\mu \psi + 2im \bar{\psi}\psi\right) \\
& \equiv \int_M \mathrm{d}^{N+1}x \mathcal{L}, \label{eq:action}
\end{aligned}
\end{equation}
where the gamma matrices $\{ \gamma^\mu \equiv e_a^\mu \gamma^a \}$ are the curved space gamma matrices which obey the Clifford algebra $\{ \gamma^\mu, \gamma^\nu \} = 2 g^{\mu \nu}$ and are related to the local flat Minkowski space gamma matrices $\{ \gamma^a \} $ which obey the flat space Clifford algebra $\{ \gamma^a,\gamma^b \} = 2 \eta^{ab}$. The Dirac adjoint is defined as $\bar{\psi} = \gamma^\dagger \gamma^0$ where $\gamma^0$ is the flat space gamma matrix. We also have $|e| = \det e^a_{\ \mu} = \sqrt{-g}$. The covariant derivative of spinors $D_\mu$ is defined via
\begin{equation}
    D_\mu \psi = \partial_\mu \psi + \Omega_\mu \psi
\end{equation}
where $\Omega_\mu$ is the spin connection related to the connection of $M$ via
\begin{equation}
    \Omega_\mu = \frac{i}{2} \Omega_{ab \mu} \Sigma^{ab}, \quad \Sigma^{ab}= \frac{i}{4}[\gamma^a,\gamma^b]
\end{equation}
and $\Omega_{ab \mu}$ are the components of the connection. For more details, see Ref.~\cite{Nakahara}.

In this study we are interested in the $(1+1)$D spacetimes, in which case the spin connection vanishes from the symmetrised action. To see this, we can substitute in the covariant derivative explicitly 
\begin{equation}
\begin{aligned}
    \mathcal{L} & = \frac{i}{2}|e| \left( \bar{\psi} \gamma^\mu D_\mu \psi - \overline{D_\mu \psi}\gamma^\mu \psi \right) \\
    & = \frac{i}{2}|e| \left( \bar{\psi} \gamma^\mu \partial_\mu \psi - \partial_\mu \bar{\psi}\gamma^\mu \psi + \bar{\psi} \{ \gamma^\mu, \Omega_\mu \} \psi \right)
\end{aligned}
\end{equation}
However, in $(1+1)$D, we have $\Omega_\mu \propto [\gamma^0,\gamma^1] \propto \gamma^3$, where $\gamma^3$ is the $(1+1)$D analogue of $\gamma^5$. As $\{ \gamma^\mu , \gamma^3 \} = 0$ for all $\mu$, then the spin connection vanishes and we arrive at the Lagrangian~\cite{Nakahara}
\begin{equation}
\mathcal{L}  =  |e| \bar{\psi} \gamma^\mu \overset{\leftrightarrow}{\partial_\mu}  \psi,
\end{equation}
where $A\overset{\leftrightarrow}{\partial_\mu}B =\frac{1}{2} \left( A \partial_\mu B - (\partial_\mu A)B \right)$.

Throughout this we assume that we are working with time-independent metrics, so that the vector $\xi = \partial_t $ is a time-like Killing vector which obeys $\mathcal{L}_\xi g = 0$. In order to perform canonical quantisation of this theory, we introduce the canonical momentum $\pi_a(x)$ of the field and apply the canonical commutation relations. We have
\begin{equation}
    \pi = \frac{\partial \mathcal{L}}{\partial \dot{\psi}} = \frac{i}{2}|e| \bar{\psi} \gamma^t
\end{equation}
The equal time canonical Poisson bracket reads
\begin{equation}
    \{ \psi_{\alpha}(t,x),\pi_{\beta}(t,y) \} = i \delta(x-y) \delta_{\alpha \beta}
\end{equation}
where the indices $\alpha,\beta$ here refer to the spinor indices. This implies that the spinor field obeys the commutation relation
\begin{equation}
    \{ \psi_\alpha(t,x),\psi_\beta^\dagger(t,y) \} = \frac{(\gamma^0 \gamma^t)_{\alpha \beta}^{-1} \delta(x-y)}{|e|}. \label{eq:curved_space_commutation_relations}
\end{equation}
Note that the factor of $1/2$ is missing despite it being present in the canonical momentum $\pi$. This is because the canonical momentum defines a constraint on phase space, as $\pi$ is linearly related to $\psi^\dagger$, which means we must employ the machinery of Dirac brackets instead of Poisson brackets to quantise this theory. It is the Dirac bracket that we upgrade to an anti-commutator using canonical quantisation. More information can be found in Ref.~\cite{Ba_ados_2016,dirac2001lectures,Srednicki_2007,barcelos1987canonical}. When taking this into account, the factor of $1/2$ vanishes.
\subsection{Relativistic limit}

\begin{figure}
    \centering
    \includegraphics[width=0.75\columnwidth]{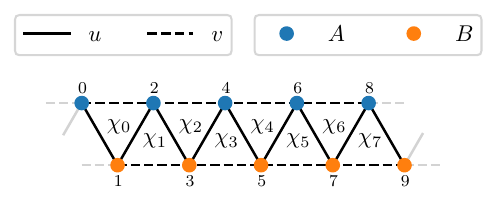}
    \caption{To reveal the relativistic behaviour, we introduce a two-site unit cell by bicolouring the lattice.}
    \label{fig:bipartite_lattice}
\end{figure}

To make the link with relativity, the lattice sites are now labelled as alternating between sub-lattices $A$ and $B$ by introducing a two-site unit cell, as shown in Fig.~\ref{fig:bipartite_lattice}. The mean field Hamiltonian of Eq.~\eqref{eq:mf_ham} can be re-parameterised as 
\begin{equation}
\begin{aligned}
H_\mathrm{MF} = \frac{1}{4} \sum_n \Big[ & -ua^\dagger_n(b_n + b_{n-1}) \\
& - \frac{iv}{4} (a_n^\dagger a_{n+1} + b^\dagger_n b_{n+1}) \Big] + \text{H.c.}, \label{eq:lattice_ham}
\end{aligned}
\end{equation}
where the fermionic modes $a_n$ and $b_n$ belong to sublattice $A$ and $B$, respectively, where now $n$ labels the unit cells. These modes obey the commutation relations $
\{ a_n,a^\dagger_m \}  = \{ b_n, b_m^\dagger\} = \delta_{nm}$, while all mixed anti-commutators vanish. The index $n$ now labels the unit cells. We Fourier transform the fermionic modes as 
\begin{equation}
a_n = \frac{1}{\sqrt{N}} \sum_{p \in \mathrm{B.Z.}} e^{i pan } a_p, \label{eq:lattice_ft}
\end{equation}
and similarly for $b_n$, where $N$ is the number of unit cells in the system, $a$ is the unit cell spacing, and $\mathrm{B.Z.} = [0,2 \pi/a)$ is the Brillouin zone and $ p = 2m\pi/L$ for $L= Na$ and integer $m$. The Fourier transformed Hamiltonian becomes
\begin{equation}
H_\mathrm{MF} = \sum_{p \in \mathrm{B.Z.}} \chi^\dagger_p h(p) \chi_p, \quad h(p) = \begin{pmatrix} g(p)  & f(p) \\ f^*(p) & g(p)  \end{pmatrix},
\end{equation}
where the two-component spinor is defined as $\chi_p = (a_p, b_p)^\mathrm{T}$ and the functions are given by
\begin{equation}
f(p) = -\frac{u}{4}(1+e^{-iap}), \quad g(p) = \frac{v}{8} \sin(ap).
\end{equation}
The dispersion relation is given by the eigenvalues of the single-particle Hamiltonian $h(p)$ which yields 
\begin{equation}
\begin{aligned}
E(p) & = g(p) \pm |f(p)| \\
& = \frac{v}{8} \sin(ap) \pm \frac{u}{4} \sqrt{2 + 2 \cos(a p) } \label{eq:app_dispersion_2}.
\end{aligned}
\end{equation}
In Fig.~\ref{fig:Chain} of the main text, it is seen that the parameter $v$ has the effect of tilting the cones as it increases and one band becomes flat at the Fermi point when $|v|/2 = |u|$.

The Fermi points $\{ p_i\}$, defined as the points for which $E(p_i) = 0$, are found at
\begin{equation}
p_0 =  \frac{\pi}{a}, \quad p_\pm  = \pm \frac{1}{a} \arccos \left( 1 -  \frac{8u^2}{v^2} \right) .
\end{equation}
The roots $p_\pm$ only exist if the argument of $\arccos$ is in the range $[-1,1]$ which implies $|v|/2 \geq |u|$ for these to appear in the dispersion. Therefore, if $|v|/2 \leq |u|$, the only Fermi point is located at $p_0 = \pi/a$ which is where the Dirac cone is located, as shown in Fig.~1 of the main text. When the cone over-tilts, so when $|v|/2 \geq |u|$, then the additional zero-energy crossings at $p_\pm$ appear which is due to the Nielsen-Ninomiya theorem which states that the number of left- and right-movers must be equal \cite{NIELSEN1981219,NIELSEN1981173}.

The continuum limit of a lattice model is an effective theory obtained by letting the lattice spacing $a \rightarrow 0$ in such a way that the Fermi velocity remains fixed. In this process, only the linear portion of the dispersion relation, near the Fermi points, is relevant as the non-linear portion of the dispersion goes off to infinite momentum. Therefore, the continuum limit is equivalent to restricting ourselves to a small neighbourhood of the Fermi points in momentum space. We outline this below. See Ref.~\cite{fradkin2013field} for more detail.

First we Taylor expand the single-particle Hamiltonian $h(p)$ about the Fermi point $p_0$ to first order in momentum which yields
\begin{equation}
h(p_0 + p)  =  \frac{1}{4}\left(u\sigma^y p -  \frac{v}{2} \mathbb{I} p \right) \equiv e_a^{\ i} \alpha^a p_i , \label{eq:h_taylor}
\end{equation}
The coefficients in the second equality are defined as $e^{\ x}_0 = - v/8,e^{\ x}_1 = u/4 $, where we have absorbed a factor of $a$ into the couplings as $au \rightarrow u$ and $av \rightarrow v$, and the Dirac matrices are $\alpha^0 = \mathbb{I},\alpha^1 = \sigma^y $. We then project the Hamiltonian of Eq.~\eqref{eq:lattice_ham} into a small region of momentum space centred on $p_0$ by truncating the summation with a cutoff $\Lambda = O(1/a)$ as
\begin{equation}
\begin{aligned}
H & \approx  \sum_{|q| < \Lambda} \chi^\dagger_{p_0 + q}  h(p_0 + q) \chi_{p_0 + q} \\
& =  \sum_{|q| < \Lambda} \chi^\dagger(q)  e_a^{\ i} \alpha^a p_i \chi(q) \label{eq:h_truncated}
\end{aligned}
\end{equation}
where we have defined the new momentum space fields $\chi(q) \equiv \chi_{p_0 + q}$, where $q$ measures the distance from the Fermi point $p_0$. 

We also truncate the discrete Fourier transform for the lattice fermions from Eq.~\eqref{eq:lattice_ft} as
\begin{equation}
\begin{aligned}
a_n & \approx \frac{1}{\sqrt{N}} \sum_{|q| < \Lambda} e^{i(p_0 + q)an}a_{p_0 + q} \\
& = e^{ip_0an} \frac{1}{\sqrt{N}} \sum_{|q| < \Lambda} e^{iqan} a(q) \\
& \equiv e^{ip_0an} a(n),
\end{aligned}
\end{equation}
which defines a slowly-varying field $a(n)$, and similarly for $b_n$ which is related to $b(n)$ analogously. We see that on the subspace near the ground state, the fermionic operators $a_n$ and $b_n$ consist of a slowly-varying field $a(n)$ and $b(n)$ respectively, with a high-frequency oscillation $e^{ip_0an}$ on top~\cite{fradkin2013field}.

We then take the limit that $a \rightarrow 0$ in such a way that the rescaled couplings $u$ and $v$ remain finite (equivalently the Fermi velocity remains fixed) and $Na = L$ remains constant. We also define $na \rightarrow x$ which we must remember is the unit cell coordinate. The cutoff $\Lambda \rightarrow \infty$ additionally, so the summation is from $\pm \infty$. Performing this limit, real space becomes a continuum and the envelope functions $a(n)$ become
\begin{equation}
    a(x) = \lim_{a \rightarrow 0} \frac{a(n)}{\sqrt{a}} = \frac{1}{\sqrt{L}} \sum_{q \in \mathrm{B.Z.}} e^{iqx} a(q) \label{eq:cont_spinor_space}
\end{equation}
and similarly for $b(x)$, where now the Brillouin zone has extended to infinity as $\mathrm{B.Z.} = [-\infty,\infty]$ with $p = 2m\pi/L$ for $m \in \mathbb{Z}$, where the re-scaling by $1/\sqrt{a}$ ensures that the limits exist and the commutation relations become continuum commutation relations. If we define the two-component spinor field $\chi(x) = (a(x),b(x))^T$, we see that this is related to the momentum space fields derived in Eq.~\eqref{eq:h_taylor} by a Fourier transform as
\begin{equation}
    \chi(q) = \frac{1}{\sqrt{L}} \int_{0}^L \mathrm{d}x e^{-iqx} \chi(x). \label{eq:cont_spinor}
\end{equation}
With this result in hand, we are now able to transform the truncated Hamiltonian of Eq.~\eqref{eq:h_truncated} back into real space, arriving at the Hamiltonian
\begin{equation}
\begin{aligned}
H &  =  \int_\mathbb{R} \mathrm{d}x \chi^\dagger(x) \left( -ie_a^{\ i} \alpha^a \overset{\leftrightarrow}{\partial_i} \right) \chi(x) \\
& \equiv \int_\mathbb{R} \mathrm{d}x \mathcal{H} , \label{eq:cont_ham}
\end{aligned}
\end{equation}
with $A\overset{\leftrightarrow}{\partial_\mu}B =\frac{1}{2} \left( A \partial_\mu B - (\partial_\mu A)B \right)$ and the Dirac $\alpha^a = (\mathbb{I},\sigma^y)$ and $\beta = \sigma^z$. We have ignored the overall factor of $1/4$ here.

This is a Hamiltonian for the slowly-varying envelope function $\chi(x)$. The corresponding action of this theory is given by
\begin{equation}
\begin{aligned}
    S & = \int d^{1+1}x  \left(i \chi^\dagger  \overset{\leftrightarrow}{\partial_t} \chi - \mathcal{H} \right)  \\
    & = \int d^{1+1}x i \bar{\chi} e_a^{\ \mu} \gamma^a \chi
\end{aligned}
\end{equation}
where we have defined $\bar{\chi} = \chi^\dagger \gamma^0$ and the gammas are related to the alpha and beta matrices via $\gamma^0 = \beta$ and $\gamma^i = \beta \alpha^i$, where $\gamma^0  = \sigma^z$ and $\gamma^1 = -i \sigma^x$ which obey the anti-commutation relations $\{ \gamma^a, \gamma^b \} = 2\eta^{ab}$, with $\eta^{ab} = \mathrm{diag}(1,-1)$. The coefficients are given by 
\begin{equation}
e_a^{\ \mu} = \begin{pmatrix} 1 & -v/8 \\ 0 & u/4 \end{pmatrix}, \quad e^a_{\ \mu} = \begin{pmatrix} 1 & v/(2u) \\ 0 & 4/u \end{pmatrix} \label{eq:tetrad}
\end{equation}
 If we assume that the couplings are upgraded to slowly-varying functions as $u \rightarrow u(x)$ and $v \rightarrow v(x)$, then the continuum limit is still a good approximation.

This action looks very similar to the action of a Dirac field on a $(1+1)$-dimensional curved spacetime, except two subtle differences. The first is that the integration measure is missing the factor of $|e|$, so it is the flat space volume element. The second is that the fields obey flat space commutation relations
\begin{equation}
\{ \chi_\alpha(x),\chi_\beta(y) \} = \delta_{\alpha \beta} \delta(x-y)
\end{equation}
which can be obtained using Eqs.~\eqref{eq:cont_spinor_space} and \eqref{eq:cont_spinor} and the fact that the component momentum space modes obey $\{ a(p),a^\dagger(q) \} = \delta_{pq}$, and similarly for $b(p)$. Therefore, the theory in its current form describes a generalised Dirac action on a flat $(1+1)$-dimensional space with space-dependent coefficients. 

In order to interpret this theory as a curved space theory, we introduce a new field
\begin{equation}
    \psi = \frac{\chi}{\sqrt{|e|}}
\end{equation}
then the fields obey the curved space commutation relations
\begin{equation}
    \{ \psi_\alpha(x),\psi_\beta(y) \} = \frac{\delta_{\alpha \beta} \delta(x-y)}{|e|}
\end{equation}
agreeing precisely with the general commutation relations of Eq.~(\ref{eq:curved_space_commutation_relations}) using the veilbein of Eq.~(\ref{eq:tetrad}). The action of Eq.~\eqref{eq:action} re-expressed in terms of the field $\psi$ is precisely the Dirac action for a spinor on a spacetime with veilbein given in Eq.~\eqref{eq:tetrad}. This veilbein corresponds to the metric
\begin{equation}
\mathrm{d}s^2 = \left( 1 - \frac{v^2}{4u^2} \right)\mathrm{d}t^2 - \frac{4v}{u^2}\mathrm{d}t \mathrm{d}x - \frac{16}{u^2}\mathrm{d}x^2.  \label{eq:app_GP_metric} 
\end{equation}

If the variables $u$ and $v$ are upgraded to slowly-varying functions of space, then the preceding calculation is still valid and the event horizon is located at the point $x_h$, where $|v(x_h)|/2 = |u(x_h)|$. In the small region in which $v$ is a slowly-varying functions of $x$, the coupling of different momentum modes will be small and can be ignored to a good approximation, leaving the diagonal terms $a^\dagger_p a_p$ only. This is quite standard to do in lattice model where the continuum is described by a Dirac equation~\cite{Sabsovich,fradkin2013field,Golan_2018}. For coordinate dependent coefficients, this is the Gullstrand-Painlev\'e metric \cite{Volovik2003} also know as the \textit{acoustic metric} which is the Schwarzschild metric of a $(1+1)$D black hole expressed in Gullstrand-Painlev\'e coordinates. This metric is referred to here as an \textit{internal metric} of the model as it depends upon the internal couplings of the Hamiltonian and not the physical geometry of the lattice. In addition, this is a fixed classical background metric and the quantum fields have no back-reaction on the metric. Quite remarkably, the phase boundary between the regions for $|v|/2 < u$ and $|v|/2 > u$ can be interpreted as the inner and outer regions of the black hole, where the phase boundary $|u| = |v|/2$  aligns with the event horizon.

In order to transform the metric of Eq.~\eqref{eq:app_GP_metric} into the Schwarzschild metric, a coordinate transformation defined as $(t,x) \mapsto (\tau,x)$ is used, where
\begin{equation}
\tau(t,x)  = t - \int_{x_0}^x \mathrm{d} z  \frac{V(z)}{U^2 - V^2(z)} ,
\end{equation}
where we have absorbed some factors into the coupling as $u/4 \rightarrow U$ and $v/8 \rightarrow V$. This maps the metric to
\begin{equation}
\mathrm{d}s^2 =  \left( 1 - \frac{V^2(x)}{U^2(x)} \right) \mathrm{d}\tau^2 -  \frac{1}{U^2(x) \left( 1 - \frac{V^2(x)}{U^2(x)} \right)} \mathrm{d}x^2, 
\end{equation}
which takes the general form of the Schwarzschild metric, where the horizon is at the location where the metric becomes singular at $U = V$ which is equivalent to $u = v/2$. 

\section{Fitting the exponent of the Laypunov vs temperature}
\begin{figure}
    \centering
    \includegraphics[width=\columnwidth]{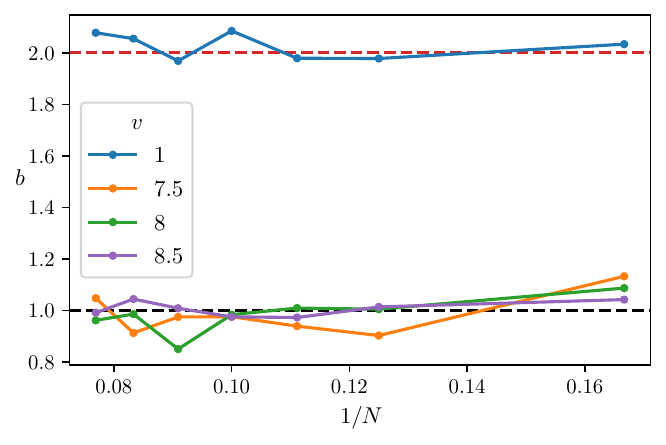}
    \caption{For different values of $v$, we fit $\lambda=(a(T^b-c)$ and extract the exponent $b$, plotting this against $1/N$. We see in XY phase ($v=1$), this value tends towards 2 (red dashed line) while in the chiral phase, $v>7.5$, this value tends towards 1 (black dashed line).}
    \label{fig:bfit}
\end{figure}

In Fig.~\ref{fig:Lyap}, we present the Lyaponov exponent vs temperature in two regimes: in the XY phase ($v=1$) where the growth is quadratic, and deep in the chiral phase ($v=8$) where the growth is shown to be linear. In this section, we explicity show that these fitting forms are suitable in the two limits by additionally parameterising the fit of $\lambda$ vs $T$ to $a(T^b - c)$ and plotting $b$ against $N$. The results are shown in Fig.~\ref{fig:bfit} where we see in XY phase, $b\approx2$, while in the chiral phase $b\approx1$, as one would hope. It is important to note that the results of the fit fluctuate dependent on the choice of $T_{\rm{min}}$ and $T_{\rm{max}}$ in the fitting window. To achieve this, different values have been taken for different $N$ and $v$, so one must be cautious during the fitting procedure. These results, nonetheless, demonstrate that with a suitable choice of fitting window, a quadratic and linear fit are reasonable to make.

\end{document}